\documentclass[%
    reprint,
    showpacs,preprintnumbers,
    showkeys,
    amsmath,amssymb,
    aps,
    pra,
    floatfix,
    longbibliography,
]{revtex4-1}

\usepackage[english]{babel}
\usepackage[utf8]{inputenc}
\usepackage{graphicx}
\usepackage{bm}
\usepackage{hyperref}
\usepackage{xcolor}
\usepackage{epsfig}
\usepackage{amsmath}
\usepackage{amsfonts}
\usepackage{amssymb}
\usepackage{amsthm}
\usepackage{tikz}
\usepackage{dsfont}
\usepackage[all]{nowidow}
\usepackage{pifont}
\usepackage{algorithm}
\usepackage{algpseudocode}
\usepackage[capitalise]{cleveref}

\usepackage[letterspace=130]{microtype}

\definecolor{refColor}{HTML}{0376E9}
\definecolor{figColor}{HTML}{E90303}
\definecolor{urlColor}{HTML}{0376E9}

\hypersetup{
    unicode=false,                 
    pdftoolbar=true,               
    pdfmenubar=true,               
    pdffitwindow=false,            
    pdfstartview={FitH},           
    pdftitle={},                   
    pdfauthor={Nicolai Lang},      
    pdfsubject={Quantum physics},  
    pdfcreator={Nicolai Lang},     
    pdfproducer={Nicolai Lang},    
    pdfkeywords={},                
    pdfnewwindow=true,             
    colorlinks=true,               
    linkcolor=figColor,            
    citecolor=refColor,            
    filecolor=magenta,             
    urlcolor=urlColor              
}

\newcommand{\Bra}[1]{\left<#1\right|}
\newcommand{\Ket}[1]{\left|#1\right>}
\newcommand{\bra}[1]{\mathinner{\langle{#1}|}}
\newcommand{\ket}[1]{\mathinner{|{#1}\rangle}}

\renewcommand{\vec}[1]{\mathbf{#1}}

\renewcommand{\vec}[1]{\boldsymbol{#1}}

\newcommand{\tr}[1]{\operatorname{Tr}\left[#1\right]}

\newcommand{\Tr}[2]{\operatorname{Tr}_{#2}\left[#1\right]}

\newcommand{\M}{\mathcal{M}}

\newcommand{\com}[2]{\left[#1,#2\right]}
\newcommand{\acom}[2]{\left\{#1,#2\right\}}

\renewcommand{\ol}{\overline}

\newcommand{\const}{\textrm{const}}

\renewcommand{\O}{\mathcal{O}}

\newcommand{\X}{\mathcal{X}}

\newcommand{\CaptionMark}[1]{\textit{#1}}

\begin{document}

\title{Entanglement Transition in the Projective Transverse Field Ising Model}

\author{Nicolai Lang}
\email{nicolai@itp3.uni-stuttgart.de}
\author{Hans Peter B\"uchler}
\affiliation{%
    Institute for Theoretical Physics III 
    and Center for Integrated Quantum Science and Technology,\\
    University of Stuttgart, 70550 Stuttgart, Germany
}

\date{\today}


\begin{abstract}
    Discrete quantum trajectories of systems under random unitary gates and
    projective measurements have been shown to feature transitions in the
    entanglement scaling that are not encoded in the density matrix.
    In this paper, we study the \emph{projective transverse field Ising model},
    a stochastic model with two noncommuting projective measurements and no
    unitary dynamics. We numerically demonstrate that their competition drives
    an entanglement transition between two distinct steady states that both
    exhibit area law entanglement, and introduce a classical but nonlocal
    model that captures the entanglement dynamics completely. Exploiting a map
    to bond percolation, we argue that the critical system in one dimension is
    described by a conformal field theory, and derive the universal scaling of
    the entanglement entropy and the critical exponent for the scaling of the
    mutual information of two spins exactly. We conclude with an interpretation
    of the entanglement transition in the context of quantum error correction.
\end{abstract}



\maketitle

\section{Introduction}

Entanglement has emerged as a powerful tool to characterize states of
matter, such as ground states of quantum systems~\cite{Eisert2010},
and access their topological properties~\cite{Levin2006,Kitaev2006}, but
also to distinguish between generic thermal states and the phenomenon of
many-body localization~\cite{Abanin2019}. Recently, a unique transition in
random quantum circuits has been identified where the entanglement entropy
of the wave function is the key observable that characterizes two different
steady states~\cite{Li2018,Chan2019,Skinner2019,Li2019,Bao2020}. This
entanglement transition is driven by the competition between random unitary
operations and projective measurements applied at discrete time steps on
the wave function. Remarkably, the density matrix of both steady states
is maximally mixed, and the transition is only visible in the average
of particular properties of wave functions over ensembles of quantum
trajectories~(see Appendix~\ref{app:transition}).
In this paper, we study a new type of entanglement transition of quantum
circuits that is driven by random projective measurements only.

Quantum circuits are an example of quantum dynamical maps where quantum
operations are applied on qubits at discrete time steps. Random
unitary operations between neighboring qubits spread entanglement
and, if this is the dominating process, entail a volume law for the
entanglement entropy~\cite{Nahum2017,Keyserlingk2018,Nahum2018}. (Note
that randomness in the unitary dynamics is not necessary to
proliferate entanglement and drive entanglement transitions; see, e.g.,
Refs.~\cite{Li2019,Tang2020,Alberton2020}.) By contrast, random projective
measurements of local observables remove entanglement from the system,
and eventually lead to wave functions with area law entanglement.
It is the competition between these two processes that gives rise
to the entanglement transition at a finite critical rate of the two
processes~\cite{Li2018,Chan2019,Skinner2019,Li2019,Bao2020}. While the
numerical simulation of generic quantum circuits is a computationally
hard problem, it was noticed that unitary gates restricted to the
Clifford group allow the time evolution of the quantum circuit
to be studied numerically even for large systems in the stabilizer
formalism~\cite{Gottesman1996,Gottesman1997,Aaronson2004,Fattal2004} (the
formal statement is referred to as \emph{Gottesmann-Knill theorem}). This
approach allowed for the precise characterization of the critical properties of
the entanglement transition~\cite{Zabalo2020}. But also analytical methods have
been contrived to unravel the nature of the transition~\cite{Bao2020,Jian2020},
such as descriptions in terms of conformal field theories~\cite{Li2020}.
However, it is well established that phase transitions with conventional
symmetry breaking, characterized by an order parameter, as well as topological
order in nonequilibrium steady states, can appear in driven quantum systems
with competing dissipative processes~\cite{Diehl2008,Lang2015}. This motivates
the question whether entanglement transitions can appear in random quantum
circuits with projective measurements only. A necessary ingredient is
certainly noncommuting, competing measurements.

In this paper, we present a detailed study of an entanglement transition
between two steady states (characterized by quantum jump trajectories of wave
functions) that both feature area law entanglement. The quantum circuit is
constructed from two noncommuting projective measurements that are applied
randomly; this model can be viewed as the natural translation of the transverse
field Ising model into a circuit of projective measurements only, and is
therefore referred to as the \emph{projective transverse field Ising model}
(PTIM).
Also in this particular case, the transition is only visible in the average
of certain entanglement measures of wave functions over ensembles of quantum
trajectories. While the quantum circuit can be studied numerically using
the stabilizer framework, we demonstrate that the entanglement dynamics
can be mapped onto a simpler, classical model which can be simulated more
efficiently. In addition, this mapping provides a simple intuition for the
spreading of entanglement in our model. We find that the transition exhibits
a behavior similar to conventional second-order quantum phase transitions,
where at the critical point long-range order is established. Here,
the role of long-range order is played by a finite mutual information
between two separated spins. Furthermore, the entanglement entropy
diverges logarithmically at the transition. These quantities unveil
two universal properties of the transition: the prefactor $\tilde{c}$
of the logarithmically diverging entanglement entropy and the critical
exponent $\kappa$ of the algebraic decay of the mutual information. We
demonstrate that the critical point is described by bond percolation and
determine $\tilde{c}=3\sqrt{3}\ln(2)/(2\pi)$ and $\kappa=2/3$ exactly for a
one-dimensional setup by mapping to a conformal field theory. Remarkably, the
prefactor $\tilde{c}$ is \emph{not} the conformal charge of the underlying
conformal field theory, which one would expect for \emph{ground states}
of critical one-dimensional systems~\cite{Calabrese2004,Calabrese2009}.

The entanglement transition studied in this paper is tightly related to quantum
error correction of topologically protected qubits encoded in the ground
state of a Majorana chain~\cite{Kitaev2001,Bravyi2010a}: the two projective
measurements can be interpreted as syndrome measurements and local errors,
respectively. In this context, the steady state characterized by finite
mutual information corresponds to the regime where the quantum information
of the code space is preserved. By contrast, in the error-dominated regime
the quantum information is lost and the mutual information vanishes. While in
the context of (active) quantum error correction it is well established that
such phase transitions exist, it is remarkable that an entanglement transition
appears even if the syndrome measurements are not recorded and no active error
correction takes place. The encoded information is still contained in the
wave function of the quantum trajectory but it would require an exponential
number of measurements to extract this information. Thus, the entanglement
transition is in general hidden from our experimental observations.

Finally, we would like to point out that, during finalizing
our manuscript, we became aware of recent, closely related
studies~\cite{Nahum2020,Lavasani2020,Sang2020,Ippoliti2020} where also
purely measurement-driven entanglement transitions between different
steady states with area law entanglement were observed. In particular,
Ref.~\cite{Nahum2020} studied the critical regime of an equivalent model in
one dimension in the Majorana representation; their findings for the critical
universal entanglement scaling $\tilde{c}$ agree with ours.

The paper is organized as follows. We introduce the quantum circuit with
two noncommuting projective measurements in \cref{sec:model}. Although
this model can be efficiently simulated as a stabilizer circuit, we
present an exact mapping to a simpler, classical model which describes the
entanglement dynamics of the system completely. The details of this mapping
are presented in \cref{sec:ccm} and the relevant observables are discussed in
\cref{sec:obs}. In \cref{sec:num} we present numerical results: we focus on
the one-dimensional chain, demonstrate the divergence of the entanglement
entropy at the critical point, and determine the prefactor $\tilde c$
of this logarithmic divergence. Then we discuss the mutual information
as an indicator of long-range entanglement and show that it exhibits the
characteristic behavior of a second-order phase transition. We determine
its critical exponent $\kappa$ and provide an intuitive interpretation
of the transition in terms of Bell clusters. The mapping to percolation
is shown in \cref{sec:perc}, which allows us to determine the position
of the entanglement transition exactly. We conclude this section with a
comparison of numerically determined critical points of the entanglement
transition for different lattices in two dimensions with known results for
bond percolation. In \cref{sec:cft}, we exploit the equivalence of our model
and bond percolation to determine the conformal field theory that describes
the critical point of the one-dimensional system and derive $\tilde{c}$
and $\kappa$ analytically. We conclude with a discussion of the relation to
quantum error correction in \cref{sec:qec}.

\section{Model}
\label{sec:model}

\begin{figure}[tb]
    \centering
    \includegraphics[width=0.8\linewidth]{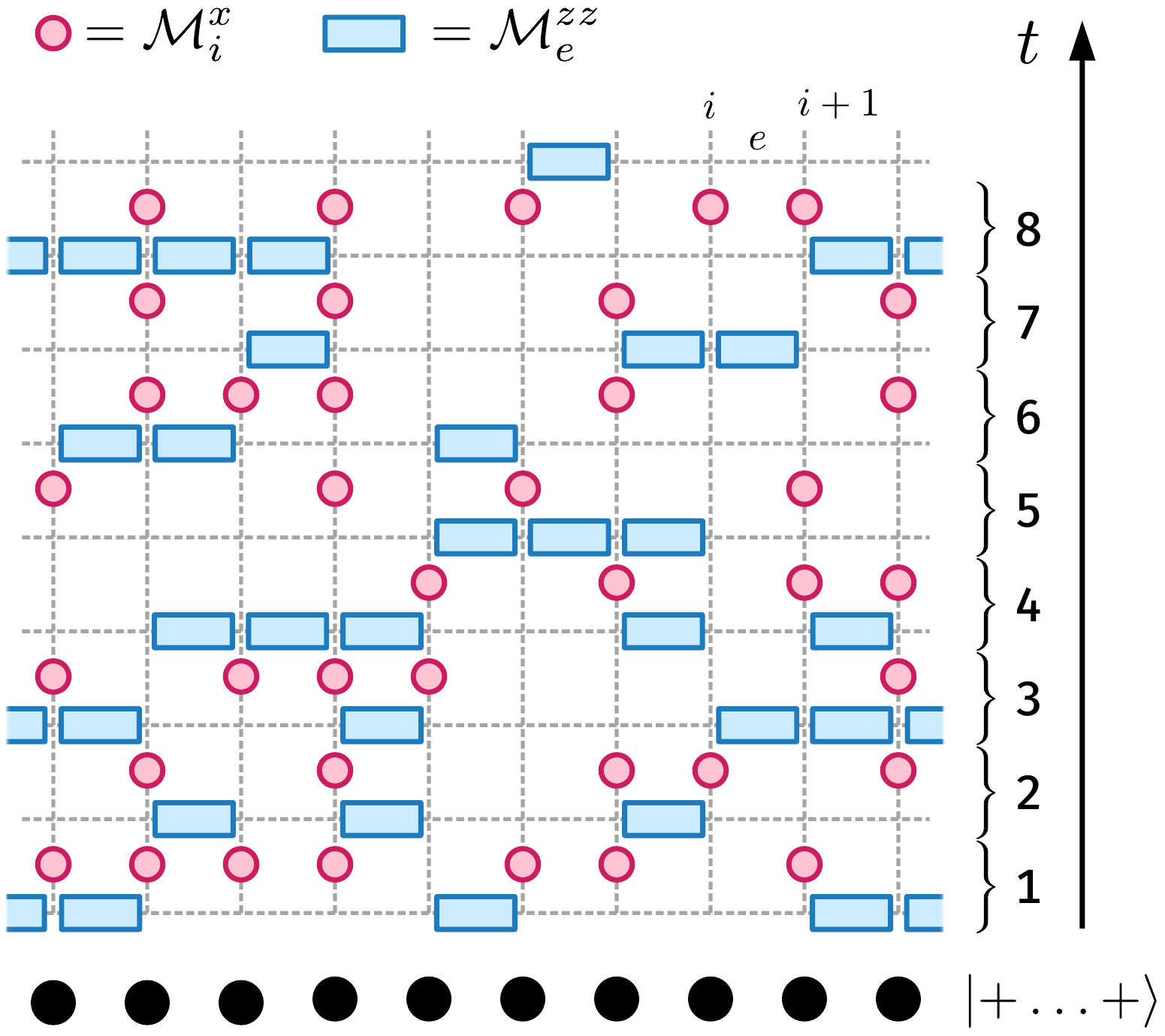}
    \caption{%
        \CaptionMark{Projective time evolution.}
        A few typical time steps for $p\approx 0.5$ on a chain of $L=10$
        spins with periodic boundaries. Blue boxes on edges denote
        measurements $\M_e^{zz}$ on adjacent spins $e=(i,i+1)$ and red
        circles measurements $\M_i^x$ on a single spin $i$. Each time step
        comprises one row of $\M_e^{zz}$ measurements followed by a row of
        $\M_i^x$ measurements. Note that the order of $\M_e^{zz}$ measurements
        does not affect the dynamics as their projectors commute. The system
        is initialized in the product state $\Ket{+\dots+}$.
    }
    \label{fig:model}
\end{figure}

We start with a detailed description of our model. Consider a
one-dimensional chain of spin-$1/2$ degrees of freedom on sites
$i\in V_L=\{1,\ldots,L\}$. Each spin is represented by Pauli matrices
$\sigma_i^\alpha$, $\alpha\in\{x,y,z\}$, and the Hilbert space is denoted as
$\mathcal{H}=\bigotimes_{i}\mathbb{C}_i^2$. The quantum circuit is defined
by projective measurements of observables $O$, and the action of such a
measurement is denoted as $\M[O]$, i.e.,
\begin{equation}
    \M[O](\Ket{\Psi})=\frac{P_\lambda\Ket{\Psi}}{\sqrt{\Bra{\Psi}P_\lambda\Ket{\Psi}}}
    \label{eq:M}
\end{equation}
is the new state after measurement of the discrete eigenvalue $\lambda$ of $O$
with probability $\operatorname{Pr}(\lambda)=\Bra{\Psi}P_\lambda\Ket{\Psi}$;
$P_\lambda$ denotes the projector onto the corresponding eigenspace. Note
that $\M[O](\Ket{\Psi})$ is a random variable with values in $\mathcal{H}$
that is parametrized by the input $\Ket{\Psi}$; $\M[O](\bullet)$ is not a
linear operator (hence the parentheses).

Throughout this paper, we are interested in measurements of the observables
$\sigma_i^x$ and $\sigma_i^z\sigma_{i+1}^z$, i.e.,
\begin{subequations}
    \label{eq:M_def}
    \begin{alignat}{3}
        \M_i^x&\equiv \M[\sigma_i^x]
        &\quad&\text{with}\quad &
        P_\lambda &= \frac{1}{2}\left(\mathds{1}+\lambda\,\sigma_i^x\right)\\
        \M_e^{zz}&\equiv \M[\sigma_i^z\sigma_{i+1}^z]
        &\quad&\text{with}\quad &
        P_\lambda &= \frac{1}{2}\left(\mathds{1}+\lambda\,\sigma_i^z\sigma_j^z\right)
    \end{alignat}
\end{subequations}
for each site $i$ and edge $e=(i,i+1)$ between adjacent sites. The
measurement results are $\lambda\in\{-1,+1\}$. It is important to point
out that $\sigma_i^x$ and $\sigma_i^z\sigma_{i+1}^z$ do not commute if they
involve the same site so that repeated measurements lead to a nontrivial
quantum dynamics.

This quantum dynamics is described as a stochastic process on $\mathcal{H}$
generated by the measurements~\eqref{eq:M_def}. In contrast to previous
studies on entanglement transitions, we do \emph{not} apply additional unitary
operations. We start with the initial product state
\begin{equation}
    \Ket{\Psi(0)}=\Ket{++\cdots+}
    \label{eq:init}
\end{equation}
with $\Ket{\pm}=\left(\Ket{0}\pm\Ket{1}\right)/\sqrt{2}$. Then, we evolve
the system iteratively as follows (\cref{fig:model}): in each time step,
we set the site variable $x_i=1$ with probability $p$ ($x_i=0$ otherwise)
and---independently---for each edge $e=(i,i+1)$, we set $z_e=1$ with
probability $1-p$ and again $z_e=0$ otherwise. The vectors $\vec x=(x_i)$
and $\vec z=(z_e)$ determine the sites (edges) on which the observables
$\sigma_i^x$ ($\sigma_i^z\sigma_{i+1}^z$) will be measured. Given the state
$\ket{\Psi(t)}$ at time $t$, the new wave function at $t+1$ is given by
(more precisely, drawn from the distribution)
\begin{equation}
    \Ket{\Psi(t+1)}=\M^x_{\vec x}\circ \M^{zz}_{\vec z} (\Ket{\Psi(t)})
    \label{eq:new}
\end{equation}
with measurements 
\begin{equation}
    \M^x_{\vec x}=\prod_{i\,:\,x_i=1} \M_i^x
    \quad\text{and}\quad
    \M^{zz}_{\vec z}=\prod_{e\,:\,z_e=1} \M_e^{zz}\,.
\end{equation}
This defines a fully projective time evolution that yields a single quantum
trajectory $\Ket{\Psi(t)}$ at discrete times $t=0,1,2,\dots$.

We are interested in characteristic properties of such wave
functions along a quantum trajectory for a given time $t$. Denote
a generic quantity as $\mathcal{X}(\Ket{\Psi(t)})$ with
$\mathcal{X}\,:\,\mathcal{H}\rightarrow\mathbb{C}$. Examples
are conventional observables such as correlations
$\mathcal{X}=\Bra{\Psi}\sigma_i^z\sigma_j^z\Ket{\Psi}$, but also the
entanglement entropy $\mathcal{X}=\mathcal{S}(A)$ of a subsystem $A\subset
V_L$. These quantities are then averaged over many different quantum
trajectories, defining the sample averages
\begin{equation}
    X\equiv\frac{1}{M}\,\sum_{\Ket{\Psi}\in\mathcal{N}}\mathcal{X}(\Ket{\Psi(t)}).
    \label{sampleaverage}
\end{equation}
Here, $\mathcal{N}=\{\Ket{\Psi(\bullet)}\}$ denotes an ensemble of $M$
randomly generated quantum trajectories.

For fixed time $t$ and $M\to\infty$, the above process defines a classical
probability distribution $\mathcal{P}(t)$ on $\mathcal{H}$. Assuming
that there exists a stationary limit, we define the \textit{projective
transverse field Ising model (PTIM)} as being characterized by
$\mathcal{P}_\infty=\lim_{t\to\infty}\mathcal{P}(t)$. Here, we are interested
in properties of $\mathcal{P}_\infty$ in dependence of the relative strength
of noncommuting measurements $p\in[0,1]$.

Note that the PTIM has a conserved symmetry: all measurements \eqref{eq:M_def}
commute with the symmetry operator $\mathcal{U}=\prod_{i\in V_L}
\sigma_{i}^{x}$. In particular, the initial state \eqref{eq:init} is
$\mathcal{U}$ invariant; this property is conserved under the projective
dynamics and restricts the accessible part of the Hilbert space. As a
consequence, the \emph{density matrix} describing the steady state of the
PTIM is only maximally mixed up to this symmetry constraint.

\section{Colored cluster model}
\label{sec:ccm}

To study the properties of the PTIM (projective transverse field Ising
model), we start with a discussion of the numerical approach that we use to
generate samples $\mathcal{N}$ of quantum trajectories $\Ket{\Psi(t)}$. It
is important to point out that measurements of Pauli operators can be
described in the stabilizer formalism~\cite{Gottesman1996,Gottesman1997}
[this is also true for the initial state~\eqref{eq:init}]; the projective time
evolution of the quantum trajectories can therefore be efficiently simulated
on a classical computer---despite the exponentially growing dimension of
$\mathcal{H}$~\cite{Aaronson2004,Fattal2004}. Although simulations in the
stabilizer formalism are reasonably efficient, generic, and well understood
to bootstrap trustworthy results, it is not the most efficient approach to
study the PTIM.

The numerical approach we leverage in this paper is based on an equivalent
\textit{classical} process that can be sampled more efficiently. This process
turns out to be \textit{nonlocal} and, in addition, provides an intuitive
picture of the mechanism that drives the PTIM entanglement transition (see
below). The derivation of this process exploits the special structure of
the PTIM and is based on the following observations.
\begin{itemize}

    \item 
        Measuring $\sigma_1^z\sigma_2^z$ in the product state (we omit
        normalizing factors)
        \begin{equation}
            \Ket{++}=\Ket{00}+\Ket{01}+\Ket{10}+\Ket{11}
            \label{eq:ex1}
        \end{equation}
        yields the entangled Bell pairs $\Ket{00}+\Ket{11}$ or
        $\Ket{01}+\Ket{10}$ with each 50\% probability. 
        We refer to both states as a two-qubit \textit{Bell cluster} (as they
        are equivalent under local unitary operations, they are identical
        from an entanglement point of view).

    \item
        Measuring $\sigma_2^z\sigma_3^z$ in the product state
        \begin{equation}
            \begin{split}
                &(\alpha\Ket{00}+\beta\Ket{11})\otimes\Ket{+}\\
                =&\alpha(\Ket{000}+\Ket{001})+\beta(\Ket{110}+\Ket{111})
            \end{split}
            \label{eq:ex2}
        \end{equation}
        yields the entangled states $\alpha\Ket{000}+\beta\Ket{111}$ or
        $\alpha\Ket{001}+\beta\Ket{110}$ with each 50\% probability. 
        The result is therefore an enlarged (three-qubit) Bell cluster.
        Note that the amplitudes survive.

    \item 
        Measuring $\sigma_1^x$ in the three-qubit Bell cluster
        \begin{equation}
            \begin{split}
                &\alpha\Ket{000}+\beta\Ket{111}\\
                =&\alpha(\Ket{+00}+\Ket{-00})+\beta(\Ket{+11}-\Ket{-11})
            \end{split}
            \label{eq:ex3}
        \end{equation}
        yields either $\Ket{+}\otimes(\alpha\Ket{00}+\beta\Ket{11})$ or 
        $\Ket{-}\otimes(\alpha\Ket{00}-\beta\Ket{11})$ with 50\% probability.
        The result is therefore a shrunken (two-qubit) Bell cluster. Note that
        again the amplitudes survive (up to a sign that depends on the
        measurement outcome).

    \item
        Measuring $\sigma_2^z\sigma_3^z$ in a state with two two-qubit Bell
        clusters
        \begin{equation}
            \begin{split}
                &(\Ket{00}+\Ket{11})\otimes (\alpha\Ket{01}+\beta\Ket{10})\\
                =&\alpha\Ket{0001}+\beta\Ket{0010}+\alpha\Ket{1101}+\beta\Ket{1110}
            \end{split}
            \label{eq:ex4}
        \end{equation}
        yields the entangled states $\alpha\Ket{0001}+\beta\Ket{1110}$ or
        $\alpha\Ket{1101}+\beta\Ket{0010}$ with each 50\% probability.
	    The result is therefore a merged four-qubit Bell cluster. Again
	    the amplitudes survive.

\end{itemize}

\begin{figure}[tb]
    \centering
    \includegraphics[width=0.9\linewidth]{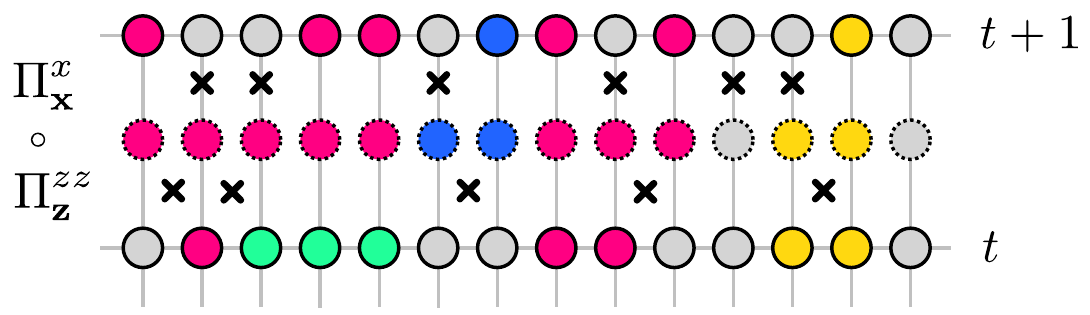}
    \caption{%
        \CaptionMark{Colored cluster model.}
        Single time step of the colored cluster model, split into two substeps:
        (1) the application of $\Pi_{\vec z}^{zz}$ on edges (crosses on faces in
        space-time) that merges/grows/nucleates clusters and (2) $\Pi_{\vec x}^{x}$
        on sites (crosses on vertical edges in space-time) that erodes clusters.
        Sites with dashed boundary represent the intermediate state between $t$
        and $t+1$ and gray sites denote single-site clusters.
    }
    \label{fig:cluster}
\end{figure}

We conclude that the dynamics of the PTIM is essentially characterized by the
nucleation, growth, decay, and merging of Bell clusters, while phase coherence is
preserved. This cluster dynamics does not depend on the specific realization of
Bell clusters (i.e., their amplitudes and their spin patterns), for instance,
\begin{equation}
    \begin{split}
        &\Ket{-000}+\Ket{-111}\,,\\
        &\Ket{+000}-\Ket{+111}\,,\\
        &\Ket{-010}+\Ket{-101}\,,\\
        &\Ket{+011}+\Ket{+100}\,.
    \end{split}
    \label{eq:example}
\end{equation}
All give rise to the same entanglement dynamics and represent the same
entanglement structure. This motivates the following classical stochastic
process (\cref{fig:cluster}).

\begin{itemize}

    \item
        The states of the system are vectors $\vec s\in\mathbb{N}_0^{L}$,
        so that the state of each site $i$ is described by a nonnegative
        integer $s_i\in\mathbb{N}_0$; $s_i=0$ encodes that site $i$ is in a
        product state and unentangled with the rest of the system. $s_i=n>0$
        marks a site that belongs to a cluster of at least two spins with
        label $n$. For example, all four states in~\eqref{eq:example} can
        be described collectively by $\vec s=(0,1,1,1)$, where $n=1$ is the
        label of the only (three-qubit) Bell cluster.

    \item
        The initial state of the process is $\vec s(0)=(0,\dots,0)$,
        corresponding to $\Ket{\Psi(0)}=\Ket{+\dots+}$ in \eqref{eq:init}
        [note that, e.g., $\Ket{\Psi(0)}=\Ket{-+--\dots+}$ would not alter
        the entanglement dynamics and therefore corresponds to the same
        state $\vec s(0)$].

    \item 
        Instead of measurements, the transformation
        \begin{equation}
            \vec s(t+1)=\Pi_{\vec x}^x\circ\Pi_{\vec z}^{zz}(\vec s(t))
            \label{eq:new_s}
        \end{equation}
        is applied iteratively with
        \begin{equation}
            \Pi_{\vec x}^x=\prod_{i\,:\,x_i=1} \Pi_i^{x}
            \quad\text{and}\quad
            \Pi_{\vec z}^{zz}=\prod_{e\,:\,z_e=1} \Pi_e^{zz}\,.
        \end{equation}
        The function $\Pi_i^x$ acts locally and is defined by $\vec
        s'=\Pi_i^x(\vec s)$ with $s_j'=s_j$ for all $j\neq i$ and $s'_i=0$.
        By contrast, the function $\Pi_e^{zz}$ for $e=(i,j)$ acts
        \emph{nonlocally} and is defined via $\vec s'=\Pi_e^{zz}(\vec s)$
        as follows (sites that are not mentioned remain unchanged).
        \begin{itemize}
            \item\textit{Case 1:} $s_i=0$ and $s_j=0$; then
                $s_i':=\operatorname{next}(\vec s)=:s_j'$. Here,
                $\operatorname{next}(\vec s)=\min(n\in\mathbb{N}\setminus
                \vec s)$ returns the smallest integer that is not used as
                a cluster label in $\vec s$. This process creates a new,
                independent cluster with two spins.
            \item\textit{Case 2a:} $s_i\neq 0$ and $s_j=0$; then $s_j':=s_i$.
                This process joins site $j$ to the cluster of site $i$.
            \item\textit{Case 2b:} $s_i=0$ and $s_j\neq 0$; then $s_i':=s_j$.
                This process joins site $i$ to the cluster of site $j$.
            \item\textit{Case 3:} $s_i\neq 0$ and $s_j\neq 0$. Let
                $s=\min(s_i,s_j)$; then set $s_l:=s$ for all sites $l$ with
                $s_l=s_i$ or $s_l=s_j$. This process merges two clusters and
                thereby reduces the number of independent clusters by one
                without reducing the number of spins that belong to clusters.
        \end{itemize}
        The last case defines a nonlocal transformation---a consequence of
        the quantumness of the PTIM where the nonlocality of clusters is
        naturally realized by entanglement.
    
\end{itemize}

Note that one can interpret the PTIM as a \textit{local quantum} simulator
for this \textit{nonlocal classical} process. In the following, we
color the sites $i$ according to their state $s_i$ and refer to this
model as the \textit{colored cluster model (CCM)}. It is this simpler
but equivalent model that we evolve and sample numerically. We also
cross-checked our results numerically for the PTIM using the stabilizer
formalism~\cite{Gottesman1996,Gottesman1997,Aaronson2004,Fattal2004}.

\section{Entanglement measures}
\label{sec:obs}

By simulating the CCM (colored cluster model), we lose access to
some properties of the PTIM. In particular, conventional expectation
values of the wave function along the quantum trajectory are no longer
accessible. However, the entanglement transition cannot be detected by
observables anyway as the density matrix in the steady state is maximally
mixed (up to symmetry constraints). Indeed, the appropriate quantities that
characterize the entanglement transition are the \emph{entanglement entropy}
and the \emph{mutual information}---both of which can be efficiently computed
using the CCM.

The \textit{entanglement entropy} of a subsystem $A\subset V_L$ for a wave
function along a quantum trajectory is defined as
\begin{equation}
    \mathcal{S}(A)\equiv-\tr{\rho_A\log_2\rho_A}
    \label{eq:EE}
\end{equation}
with $\rho_A=\Tr{\rho}{V_L\setminus A}$ the reduced density matrix of the
subsystem. In terms of Bell clusters, $\mathcal{S}(A)$ counts the number of
clusters with support both in $A$ and $\overline{A}\equiv V_L\setminus A$,
a quantity that can be easily computed from CCM states $\vec s(t)$. Note that
here we define the entanglement entropy with the binary logarithm $\log_2$
such that each Bell cluster contributes $1$ instead of $\ln 2$.

The actual quantity of interest is the entanglement entropy $\mathcal{S}(A)$
averaged over many, randomly sampled quantum trajectories. Therefore,
we denote by $S(A)$ the sample-averaged entanglement entropy as defined
in~\eqref{sampleaverage}. We also define $S_L(l)$ as the (sample-averaged)
entanglement entropy of $l$ contiguous spins in the center of a chain
with $L$ sites, i.e., $A$ comprises the $l$ sites in the interval
$[L/2-l/2,\dots,L/2+l/2)$.

Now consider two disjoint subsystems $A,B\subset V_L$. The \textit{mutual
information} $I(A,B)$ between $A$ and $B$ is defined as
\begin{equation}
    I(A,B)\equiv S(A)+S(B)-S(A\cup B)\,.
    \label{eq:I}
\end{equation}
A nonvanishing value of $I(A,B)$ indicates entanglement between
the subsystems $A$ and $B$ for wave functions along a quantum
trajectory~\cite{Casini2004,Chiara2018}.
Here we are mainly interested in the mutual information between two spins
at sites $i,j\in V_L$, that is
\begin{equation}
    I(i,j)\equiv S(\{i\})+S(\{j\})-S(\{i,j\})\,.
    \label{eq:II}
\end{equation}
For a state $\Ket{\Psi}$ along a quantum trajectory of the PTIM, $I(i,j)$
is an $L\times L$ matrix that encodes the structure of Bell clusters in the
system completely. For example, $\Ket{\Psi}=\Ket{0,0,0,0}_{1234}$ but also
$\Ket{\Psi}=(\Ket{0,0}+\Ket{1,1})_{12}\otimes(\Ket{0,1}+\Ket{1,0})_{34}$
yields $I(1,4)=0$, since spins 1 and 4 are
not part of a common Bell cluster. By contrast,
$\Ket{\Psi}=(\Ket{0,0}+\Ket{1,1})_{14}\otimes(\Ket{0,1}+\Ket{1,0})_{23}$ yields
$I(1,4)=2$ and for $\Ket{\Psi}=\Ket{0,0,0,0}_{1234}+\Ket{1,1,1,1}_{1234}$
we have $I(1,4)=1$; in both cases, spins 1 and 4 belong to the same Bell
cluster. If we consider $I(i,j)$ as the adjacency matrix of a graph, the
connected components of this graph are in one-to-one correspondence with the
Bell clusters of $\Ket{\Psi}$. In the language of the CCM, it is $I(i,j)=0$
for sites of different color $s_i\neq s_j$ (or $s_i=0=s_j$) and $I(i,j)=1$
for sites of the same color $s_i=s_j\neq 0$; for clusters that contain only
the two sites $i$ and $j$, it is $I(i,j)=2$.

\section{Numerical results}
\label{sec:num}

For the following results, we sampled typically $M\sim 10^5-10^7$ trajectories
for sufficiently long time $t\sim 2\times 10^3$ such that all quantities of
interest reached their equilibrium values in the steady state. The quantities
of interest are the sample-averaged entanglement entropy $S_L(l)$ and the
mutual information $I(i,j)$.

\begin{figure}[tb]
    \centering
    \includegraphics[width=0.9\linewidth]{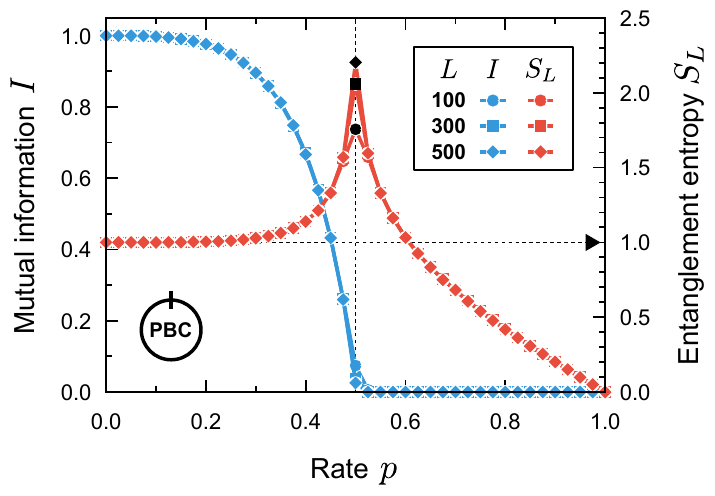}
    \caption{%
        \CaptionMark{Entanglement transition.} 
        Numerical results for the entanglement entropy $S_L(L/2)$ (red)
        and the mutual information $I(1,1+L/2)$ (blue) as functions of the
        rate $p$ for systems of length $L=100,300,500$ (circles, squares,
        diamonds) with periodic boundary conditions (PBC). At the critical
        point $p=0.5=p_c$, the entanglement entropy grows logarithmically
        with the system size (black markers). Each point is based on $10^5$
        sampled trajectories.
    }
    \label{fig:1D_p}
\end{figure}

\subsection{Entanglement entropy}

In~\cref{fig:1D_p} we show the entanglement entropy $S_L(L/2)$ as a function
of $p$ for chains of lengths $L=100,300,500$ with periodic boundaries. For
$p\to 0$ (only $\M_e^{zz}$), the entanglement entropy saturates at 1 since the
system approaches a global Bell cluster $\Ket{\vec m}+\Ket{\overline{\vec
m}}$ where $\Ket{\overline{\vec m}}\equiv\mathcal{U}\Ket{\vec m}$ with
$\vec m\in\mathbb{Z}_2^L$ a random spin pattern in the $z$ basis; here
again $\mathcal{U}=\prod_i\sigma_i^x$ denotes the global symmetry which
is conserved along the quantum trajectory. For $p\to 1$ (only $\M_i^x$),
the system approaches the unentangled product state $\Ket{+\dots +}$
so that $ S_L(L/2)$ vanishes smoothly. For $0<p<1$ there seems to be a
slow divergence at the critical value $p_c \approx 0.5$ that leads to
a weakly nonanalytic behavior of $S_L(L/2)$ for $L\to\infty$. Below, we
will demonstrate analytically that the transition indeed takes place at the
critical point $p_c = 0.5$. The two regimes can be understood intuitively in
the context of Bell clusters if we recall that $S_L(L/2)$ counts the number
of independent Bell clusters with support in both halves of the system.
\begin{itemize}
    \item For $p\gg p_c$, the projections onto $\Ket{\pm}$ dominate and make
        the clusters decay rapidly; they cannot grow to extensive size and
        cross the two boundaries of the subsystem rarely.
    \item For $p\approx p_c$, nucleation and growth of clusters on one side
        and annihilation and decay on the other side are balanced. Clusters
        become deconfined and spread throughout the system. Typically,
        two independent clusters connect the two halves of the system: one
        located at each of the two boundaries of the subsystem. In rare
        cases, additional, independent clusters contribute entanglement,
        so that $S_L(L/2)\gtrsim 2$ for long chains.
    \item For $p\ll p_c$, the growth of clusters dominates. Since the probability
        that two independent clusters merge grows exponentially with their surface, the
        probability for two or more extensive clusters vanishes exponentially. This is
        a condensation mechanism where newly created clusters (``condensation nuclei'')
        quickly get absorbed by the macroscopic cluster (the ``condensate''). This
        explains why $S_L(L/2)\to1$ quickly saturates.
\end{itemize}
Thus the average lifetime of newly spawned clusters vanishes quickly for
both $p\to 1$ and $p\to 0$. For $p\to 1$, this is due to the destructive
force of the $\M_i^x$ measurements that dissolve the clusters. For $p\to 0$,
the clusters do not dissolve but get absorbed by the condensate and this
mechanism becomes more efficient when the density of the condensate increases.

\begin{figure}[tb]
    \centering
    \includegraphics[width=1.0\linewidth]{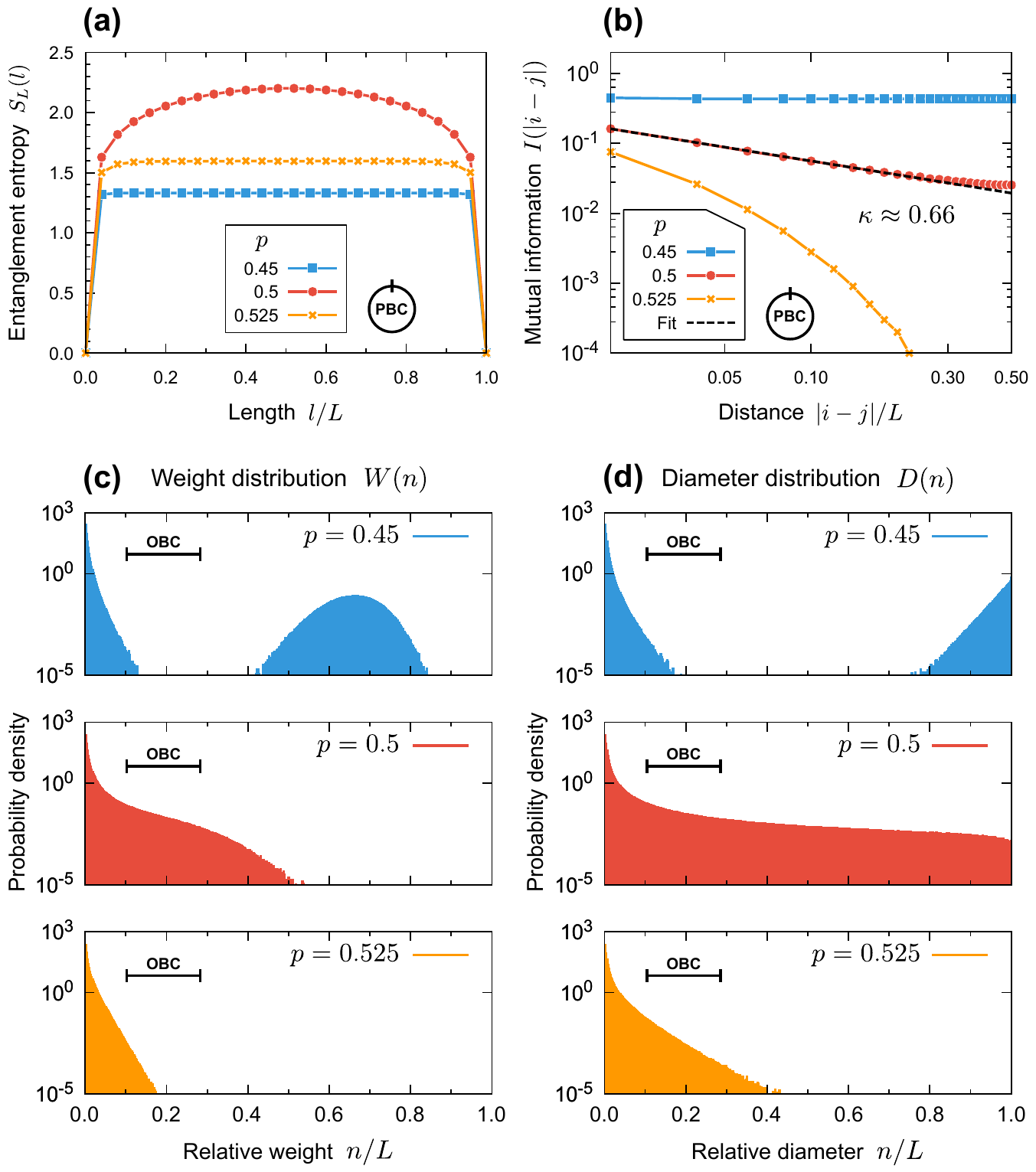}
    \caption{%
        \CaptionMark{Entanglement structure.}
        Various quantities characterizing the entanglement structure of a chain
        with length $L=500$ above, at, and below the critical point; all results
        are based on $10^6$ sampled trajectories:
        (a)~entanglement entropy $S_L(l)$ as a function of the length $l/L$ of
        the subsystem for a chain with periodic boundary conditions. The system
        obeys an area law in both phases, $p=0.45<p_c$ and $p=0.525>p_c$,
        with a logarithmic contribution at the critical point $p=0.5=p_c$.
        (b)~Mutual information $I(i,j)=I(|i-j|)$ as a function of the distance
        $|i-j|/L$ for a chain with periodic boundary conditions. $I(|i-j|)$
        vanishes exponentially for $p>p_c$ and saturates at a finite value
        for $p<p_c$. At the critical point $p=p_c$, it is well described by
        the algebraic decay $I(|i-j|)=\alpha\,|i-j|^{-\kappa}$ with fitted
        exponent $\kappa\approx 0.66$ and nonuniversal fit parameter $\alpha$
        (dashed black line).
        (c)~Probability density $W(n)\times L$ of the relative weight $n/L$
        of Bell clusters for a chain with open boundary conditions. At the
        entanglement transition, clusters of extensive weight emerge.
        (d)~Probability density $D(n)\times L$ of the relative diameter $n/L$
        of Bell clusters for a chain with open boundary conditions. At the
        entanglement transition, clusters with diameters on all length
        scales exist.
    }
    \label{fig:1D_l}
\end{figure}

In~\cref{fig:1D_l}(a) we show the behavior of $S_L(l)$ for $0\leq
l\leq L$ for different parameters $p$. As expected, we observe that
for $p\neq p_c$ the entanglement entropy saturates quickly, indicating
area law entanglement. However, this behavior is modified at the critical
point by a logarithmic contribution. The slow divergence of $S_L(l)$ at
criticality for $L\to\infty$ with $l/L=\const$ (recall \cref{fig:1D_p}),
and for $l\to\infty$ with $l\ll L$ is a well-known feature of critical
systems in one dimension that can be described by a conformal field theory:
the scaling law for ground states of conformally invariant systems with
periodic boundaries at the critical point is asymptotically described
by~\cite{Calabrese2004,Calabrese2009}
\begin{equation}
    \begin{split}
        S_L(l)
        \sim\frac{c}{3}\log_2\left[\frac{L}{\pi}\sin\left(\pi\frac{l}{L}\right)\right]
        \,\stackrel{l\ll L}{\approx}\,\frac{c}{3}\log_2(l)
    \end{split}
    \label{eq:s_cft}
\end{equation}
with the central charge $c$ (up to a nonuniversal constant).

\begin{figure}[tb]
    \centering
    \includegraphics[width=0.8\linewidth]{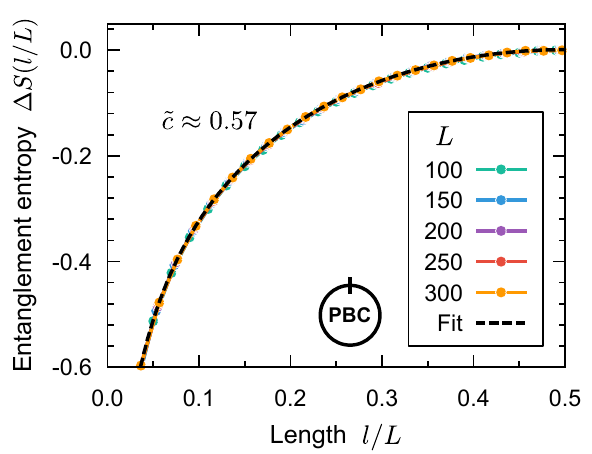}
    \caption{%
        \CaptionMark{Critical scaling.}
        Entanglement entropy $\Delta S(l/L)$ as a function of the subsystem
        size $l/L$ for chains of different length $L$ with periodic boundary
        conditions at the critical point $p=0.5=p_c$. The data collapse is
        almost perfect and the prediction for critical systems of the form
        $\Delta S(l/L)=\tilde c/3\log_2\xi(l/L)$ with $\xi(x)=\sin(\pi x)$
        describes the dependency remarkably well for the fit parameter $\tilde
        c\approx 0.57$ (dashed black line). Each curve is based on $10^6$
        sampled trajectories.
    }
    \label{fig:S}
\end{figure}

In the following, we analyze whether the critical entanglement scaling of
the PTIM exhibits the same behavior. To this end, we consider the normalized
entanglement entropy
\begin{equation}
    \Delta S(l/L)\equiv S_L(l)- S_L(L/2)\sim
    \frac{\tilde{c}}{3}\log_2\left[\sin\left(\pi\frac{l}{L}\right)\right] 
    \label{eq:s2_cft}
\end{equation}
that is expected to be independent of the system size~$L$. In~\cref{fig:S} we
plot this quantity for system sizes $L=100,\dots,300$ at criticality. The
collapse of data for different $L$ is remarkable and, fitting
Eq.~\eqref{eq:s2_cft} to our numerical data, we find the prefactor $\tilde
c\approx 0.57$. This observation suggests that the critical properties of the
PTIM are described by a conformal field theory. Indeed, we will argue below
that the critical properties are determined by a conformal field theory with
central charge $c=0$, while the prefactor of the entanglement entropy takes
the exact value $\tilde c=3\sqrt{3}\ln(2)/(2\pi)\approx 0.573$. In contrast
to ground states of critical one-dimensional quantum systems, the prefactor
of the entanglement entropy is not the central charge.


\subsection{Mutual information}

Next, we analyze the sample-averaged mutual information $I(i,j)$ in the
steady state. A finite value indicates a finite probability to find a Bell
cluster that encompasses sites $i$ and $j$. In~\cref{fig:1D_p}, the mutual
information $I=I(1,1+L/2)$ as a function of $p$ for a chain with periodic
boundaries is shown. Clearly the system undergoes a continuous entanglement
transition at the critical value $p_c\approx 0.5$, where $I>0$ for $p<p_c$
and $I=0$ for $p>p_c$. This is consistent with our interpretation above:
below the critical value, there is a ``condensate'' of Bell clusters,
that is, a single, macroscopic cluster that permeates the whole system,
creating entanglement between spins that are far apart. The limit $I_\infty =
\lim_{|i-j|\to\infty} I(i,j)$ is a measure for the density of the macroscopic
cluster and continuously converges to 1 for $p\to 0$. This behavior is in
close analogy to a conventional second-order quantum phase transition where,
at the critical point, long-range order is established. Similarly, at the
entanglement transition, the mutual information for distant spins attains
a finite value.

In~\cref{fig:1D_l}(b) we show the behavior of $I(i,j)$ for increasing
distance between the two spins [note that $I(i,j)=I(|i-j|)$ because of
translation invariance]. While for $p>p_c$ the mutual information $I(i,j)$
vanishes exponentially with the distance, it saturates at a finite value
for $p<p_c$. However, at the critical point $p_c$, it exhibits an algebraic
decay with critical exponent $\kappa$,
\begin{equation}
    I(i,j)=I(|i-j|)\sim\frac{\alpha}{|i-j|^\kappa}\,,
\end{equation}
with nonuniversal parameter $\alpha$. From the numerical results, we
find the critical exponent $\kappa\approx 0.66$. Below, we determine this
critical exponent by arguments based on the mapping of the critical regime
to a conformal field theory and find the exact value $\kappa=2/3$.

\subsection{Distribution of Bell clusters}

To quantify the emergence of a macroscopic cluster at the entanglement
transition, we define the \textit{diameter} $d(B)$ of a Bell cluster $B\subset
V_L$ as
\begin{equation}
    d(B)=\max\{|i-j|\,|\,i,j\in B\}
\end{equation}
for systems with \emph{open} boundaries; this is just
$|i_\leftarrow-i_\rightarrow|$ with the leftmost (rightmost) spin
$i_\leftarrow$ ($i_\rightarrow$) that belongs to $B$. In addition, we define
the \textit{weight} $|B|$ as the number of spins that make up the cluster $B$.
Let $\mathcal{B}_\Psi$ denote the set of all Bell clusters of a given
wave function $\Ket{\Psi}$ along a quantum trajectory (we count single
spins as one-qubit Bell clusters). The distribution $\mathcal{D}(n)$ with
$n=0,1,\dots,L-1$ for this wave function is defined as
\begin{equation}
    \mathcal{D}(n)=\left|\{B\in\mathcal{B}_\Psi\,|\,d(B)=n\}\right|\,/\,|\mathcal{B}_\Psi|\,,
\end{equation}
and similarly for the weight
\begin{equation}
    \mathcal{W}(n)=\left|\{B\in\mathcal{B}_\Psi\,|\,|B|=n\}\right|\,/\,|\mathcal{B}_\Psi|\,.
\end{equation}

In~\cref{fig:1D_l}(d), we show the diameter distribution $D(n)$, i.e., the
average over quantum trajectories of the distribution $\mathcal{D}(n)$,
as a function of the relative diameter $n/L$. In~\cref{fig:1D_l}(c), we
show the averaged distribution $W(n)$ as a function of the relative weight
$n/L$. Note that, at the entanglement transition, the distribution $D(n)$
features a long tail (demonstrating the existence of clusters on all length
scales), while for $p<p_c$ it becomes bimodal with considerable contributions
for diameters $n\sim L$, indicating the emergence of clusters that permeate
the system. By contrast, the weight distribution $W(n)$ evolves towards a
saddle point at $p_c$, which then leads to a second maximum at weights $n\sim
L/2$ that shifts continuously to $n\sim L$ for $p\to 0$. These observations
illustrate that the extensive cluster is sparse close to the critical point
and grows in density for $p\to 0$, where more and more spins ``condense''
into the macroscopic Bell cluster.

To conclude the discussion of numerical results, we point
out that the presence of a finite density of $z$-polarized spins
$\ket{\dots\uparrow\dots\downarrow\dots}$ in the initial state \eqref{eq:init}
(``$z$ poisoning'') alters the entanglement dynamics dramatically. In
particular, for $p\to 0$ both entanglement entropy and mutual information
vanish smoothly as no global Bell cluster can be established. Indeed,
while there is no process that can \emph{collapse} Bell clusters for purely
$x$-polarized initial states, $z$-polarized spins trigger an avalanche of
cluster collapses by $\M_e^{zz}$ measurements that (for small $p$) quickly
drive the system into a product state. Interestingly, even with $z$ poisoning,
there are still critical fluctuations at $p_c$ as indicated by a sharp peak
of the entanglement entropy. We do not consider effects of $z$ poisoning in
this paper.

\section{Percolation}
\label{sec:perc}

\begin{figure*}[tb]
    \centering
    \includegraphics[width=0.9\linewidth]{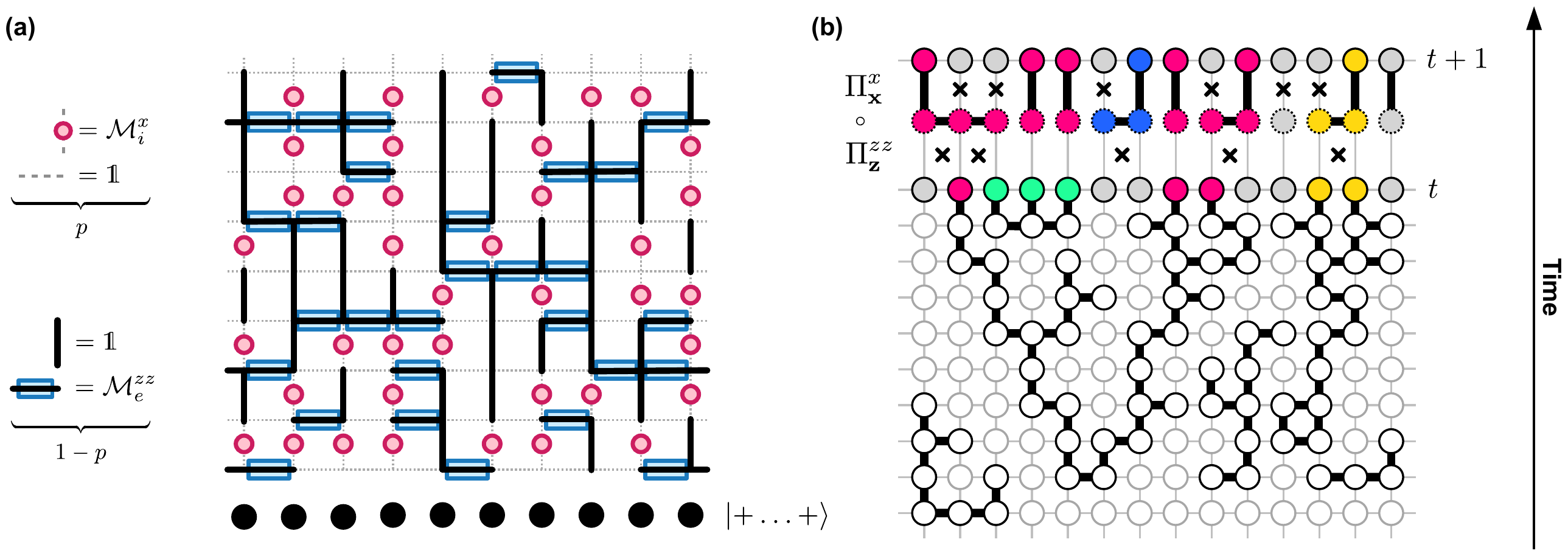}
    \caption{%
        \CaptionMark{Percolation.}
        (a)~The same measurement patter as in~\cref{fig:model}. Here we
        mark horizontal edges with $\M_e^{zz}$ measurements and vertical
        edges \textit{without} $\M_i^x$ measurements as active (bold black
        lines). Then, the probability for both horizontal and vertical
        edges to be active is $1-p$ and the projective dynamics gives rise
        to isotropic bond percolation on a square lattice in space-time.
        (b)~A possible history for the CCM state in~\cref{fig:cluster}. Sites
        at time $t$ have the same color if and only if they are connected
        via active edges in space-time.
    }
    \label{fig:perc}
\end{figure*}

The PTIM and the CCM (which captures the entanglement properties of the PTIM)
are intimately linked to \emph{bond percolation}. The mapping is illustrated
in~\cref{fig:perc}(a) and described in the following. The discrete time steps
give rise to a square lattice in space-time where each time step comprises
one horizontal row of edges and all vertical edges that connect it to the
next row. If we mark horizontal edges with $z_e=1$ and vertical edges with
$x_i=0$ as ``active'' (bold black edges), the probability for activity is
in both cases $1-p$. Thus every sequence of measurements on the spins on
a one-dimensional PTIM is in one-to-one correspondence with a pattern of
active bonds on its space-time square lattice. Following our observations
that led to the construction of the CCM, it is easy to see that, at a given
time $t$, two spins $i$ and $j$ are entangled (belong to the same Bell
cluster; in the CCM: have the same color) if and only if the two sites are
connected by a path of active edges on the space-time lattice in the past;
see~\cref{fig:perc}(b). This observation immediately implies that the critical
point for the entanglement transition of the PTIM coincides with the transition
for bond percolation; on a square lattice, this transition takes place at
$p_c=0.5$ (which follows exactly from duality arguments~\cite{Sykes1963}).
This is in agreement with the numerical results above.

\begin{table}[htb]
    \caption{%
	\CaptionMark{Critical values.} Estimates of critical values $p_c$
    for various lattices. In 2D, we compare them with numerical values
    $\tilde p_c$ from Ref.~\onlinecite{Marck1997} for three-dimensional bond
    percolation on stacks of the corresponding 2D lattices. Values marked
    with an asterisk are exact. Our estimates are based on lattices up to
    $50\times 50$ spins.
    }
    \centering
    \begin{ruledtabular}
    \begin{tabular}{cllll}
        Dimension & Lattice & $p_c$ & $\tilde p_c$ & {\scriptsize Percolation lattice} \\
        \hline
        1 & -          & $0.5$      & $0.5^*$ &{\scriptsize Square}             \\
        \hline
        2 & Square     & $0.75$    & $0.7512$ &{\scriptsize Cubic}              \\ 
        2 & Kagome     & $0.74$    & $0.7437$ &{\scriptsize stacked Kagome}     \\
        2 & Honeycomb  & $0.70$    & $0.6907$ &{\scriptsize stacked Honeycomb}  \\ 
        2 & Triangular & $0.83$    & $0.8140$ &{\scriptsize stacked Triangular} \\ 
    \end{tabular}
    \end{ruledtabular}
    \label{tab:h}
\end{table}

This close relation between bond percolation and the PTIM allows us immediately
to determine also the critical point of the entanglement transition in
higher dimensions. It is straightforward to define the PTIM on arbitrary
lattices $\mathcal{L}$ where the measurements $\M_i^x$ act with probability
$p$ on vertices $i\in V(\mathcal{L})$ and the measurements $\M_e^{zz}$ with
probability $1-p$ on edges $e\in E(\mathcal{L})$. The corresponding CCM is
then induced by bond percolation on the half-infinite stack of lattices
$\mathcal{L}$ with vertical edges connecting vertices of adjacent layers
(for instance, the PTIM on the two-dimensional square lattice is described
by bond percolation on the (2+1)-dimensional cubic lattice). We estimated
the critical values $p_c$ for square, Kagome, honeycomb, and triangular
lattice from simulations with up to $50\times 50$ spins. In~\cref{tab:h}
we compare these values with known (numerical) results for the corresponding
bond percolation problems in three dimensions~\cite{Marck1997} and find
reasonable agreement between them.

Furthermore, we would like to briefly comment on the more generic case
where $\M_i^x$ and $\M_e^{zz}$ occur with independent probabilities $p^x$
and $p^z$, respectively. Then the connectivity on the space-time lattice
is determined by \textit{anisotropic} bond percolation~\cite{Redner1979}
with probability $p_\perp=p^z$ for horizontal edges and $p_\parallel=1-p^x$
for vertical edges. In two dimensions it can be shown by duality arguments
that the system is critical for $p_\perp+p_\parallel=1$, or, equivalently,
$p^x=p^z$~\cite{Sykes1963}. If we choose the parametrization $p^z=q$ and
$p^x=rq$, the entanglement transition occurs for $r_c=1$ and is independent
of $q$ (which quantifies the overall measurement rate). We verified this
numerically and found no dependence of the entanglement transition of the PTIM
in one dimension on the measurement rate $q$. However, in higher dimensions,
the relation between the critical values $p_{\perp,c}$ and $p_{\parallel,c}$
is no longer linear~\cite{Redner1979}. As a consequence, we expect the critical
ratio $r_c=p^x_c/p^z_c$ for the PTIM on the square lattice to depend on the
measurement rate $q$. We checked this numerically and indeed found a shift
towards larger ratios $r_c$ in the limit $q\to 0$.

Finally, we point out that in the limit $q\to 0$ measurements become
rare events in each time step so that the dynamics can be approximated by
independent Poisson processes on all sites and edges for $\M^{x}_i$ and
$\M^{zz}_e$ measurements with rate parameters $\lambda^x/\lambda^z=r$.
Then, the order of measurements becomes irrelevant---in contrast
to the PTIM where in each time step we first apply $\M_e^{zz}$
and subsequently $\M_i^x$; cf.~Eq.~\eqref{eq:new}. In this limit,
the process belongs to the family of \textit{continuum random cluster
models}~\cite{Bezuidenhout1991,Grimmett2004,Grimmett2008,Grimmett2008a}
in $d+1$ dimensions which are known to describe quantum $Q$-state Potts
models~\cite{Deng2004,Grimmett2008,Grimmett2008a} in $d$ dimensions. In
particular, percolation is described by the $Q\to 1$ limit of the Potts model,
which is an important observation for the derivation of the correct conformal
field theory at the critical point of the (1+1)-dimensional PTIM.

\section{Conformal field theory at the critical point}
\label{sec:cft}

The close relation between the PTIM in one dimension and bond percolation
on a two-dimensional square lattice allows us to derive the conformal field
theory describing the critical point of the entanglement transition. First,
note that bond percolation on the square lattice is the simplest random
cluster model with cluster weight $Q=1$ (more generally, cluster models are
equivalent to classical $Q$-state Potts models~\cite{Grimmett2004}). Planar
random cluster models can be mapped to six-vertex models~\cite{Baxter1976}
which, at the critical point, have an equivalent description as a dense gas
of oriented loops with weight $\sqrt{Q}$~\cite{cardy1989}. Interpreting the
oriented loops as contour lines of a discrete ``height'' field $\phi(\vec
x)\in \pi\mathbb{Z}$ establishes an equivalent description in terms of
a solid-on-solid (SOS) model~\cite{Beijeren1977,nienhuis1987}. At large
distances and after coarse graining, the height field can be approximated
by a continuous field $\Phi(\vec x)\in\mathbb{R}$ and the solid-on-solid
model renormalizes to a Gaussian fixed point with coupling $g=1-e_0$ where
$\sqrt{Q}=2\cos(\pi e_0)$~\cite{Knops1980,Nijs1983,cardy2008}. If defined on
a cylinder (corresponding to a periodic PTIM in one dimension), the correct
weighting of noncontractible loops makes it necessary to put charges $\pm e_0$
on the two boundaries of the cylinder~\cite{Baxter1976,Bloete1986,cardy2008}
by inserting vertex operators $V_\pm^0=\exp(\pm ie_0\Phi)$. This modifies
the vacuum energy on the cylinder and shifts the central charge to
$c=1-6e_0^2/(1-e_0)$~\cite{cardy2008}. For percolation, we have $e_0=1/3$ and
the central charge vanishes, i.e., $c=0$; therefore, the prefactor $\tilde c$
for the entanglement entropy~\eqref{eq:s2_cft} must play another role.

\begin{figure}[tb]
    \centering
    \includegraphics[width=0.95\linewidth]{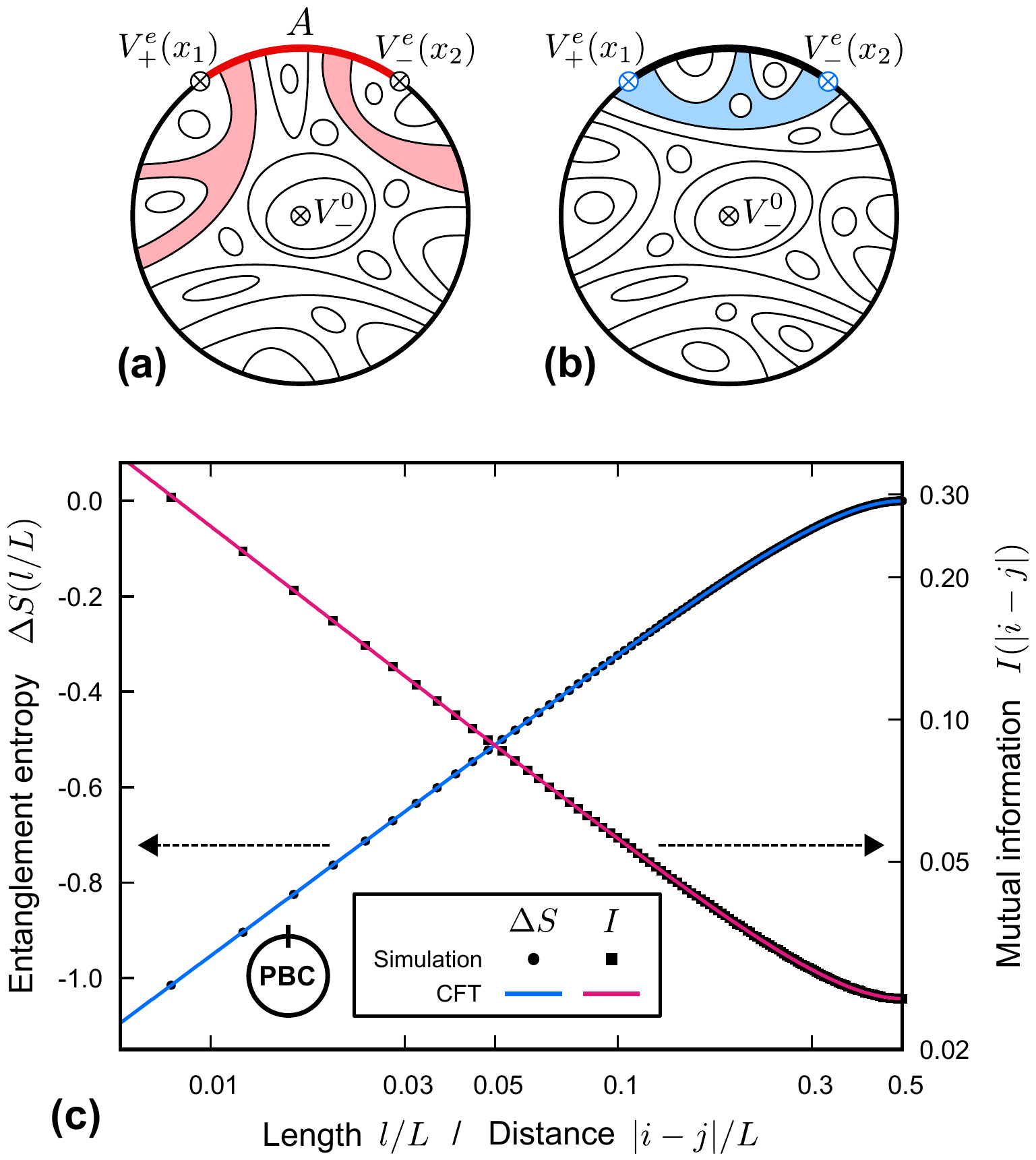}
    \caption{%
        \CaptionMark{Conformal field theory.}
        (a),(b)~Field configurations on the half-infinite cylinder mapped to the
        complex plane. Vertex operator insertions are indicated by $\otimes$.
        (a)~Scaling of the entanglement entropy $S(A)$. Field configurations with
        contours that connect $A$ (red segment) with the environment (black boundary)
        contribute to the entanglement between them (light red domains).
        (b)~Scaling of the mutual information $I(x_1,x_2)$. Field configurations
        that connect the points $x_1$ and $x_2$ (light blue domain) contribute to
        the mutual information between them.
        (c)~Comparison of numerical results (squares and bullets) and
        theoretical predictions (solid lines) for the PTIM at criticality. The
        simulations are based on $10^7$ trajectories for a system of length
        $L=500$ with periodic boundary conditions. The CFT predictions are
        of the form $\Delta S(l/L)=\tilde c/3\log_2\xi(l/L)+\alpha$ and
        $I(i,j)=I(|i-j|)=\beta\,\xi^{-\kappa}(|i-j|)+\gamma$ with $\tilde c$
        and $\kappa$ as given in the text and $\alpha$, $\beta$, $\gamma$
        nonuniversal fit parameters; $\xi(x)=\sin(\pi x)$ accounts for
        the finite size and the periodic boundaries. Note that the plot
        for $\Delta S$ is logarithmic on the $l/L$ axis; the plot for $I$
        is logarithmic on both axes.
    }
    \label{fig:cft}
\end{figure}

In the following, we derive the prefactor $\tilde{c}$ for the
asymptotic behavior of the entanglement entropy in the conformal field
theory. The approach is motivated by recent results on the valence
bond entanglement entropy in the ground state of an antiferromagnetic
spin chain~\cite{Jacobsen2008}. We start by considering a half-infinite
cylinder by shifting the lower boundary to infinity. Such a half cylinder is
conveniently mapped to the complex plane, which leaves us with a disk where
$V_-^0$ is inserted at the origin and $V_+^0$ somewhere on the boundary. We
follow now the lines of Ref.~\onlinecite{Jacobsen2008} and split $V_+^0$ into
two vertex operators $V_\pm^e=\exp[i(\pm e+e_0/2)\Phi]$ with scaling dimension
\begin{equation}
    h=\frac{e^2-(e_0/2)^2}{1-e_0}
    \label{eq:h}
\end{equation}
and $e\in\mathbb{R}$ a free parameter; see~\cref{fig:cft}(a).
The pair of vertex operators $V_\pm^e$, inserted at $x_1$ and $x_2$ on the
boundary, modifies the weight of loops that connect the boundary segment
$A=[x_1,x_2]$ of length $\Delta x=|x_1-x_2|$ with the rest of the boundary
$\overline{A}$. Therefore, the correlation function of the vertex operators can
be written as
\begin{equation}
    \mathcal{V}_A(w)=
    \langle V_+^e(x_1)V_-^e(x_2)\rangle 
    =\frac{\sum \tilde{w}^{\tilde N} w^{N_A}
    }{\sum \tilde{w}^{\tilde N+N_A}}
    \sim \frac{1}{\Delta x^{2h}}
    \label{eq:VV}
\end{equation}
where the sums go over all allowed loop configurations. $N_A$ is the number
of loops connecting $A$ and $\overline{A}$, whereas $\tilde N$ counts the loops
attached with both ends either to $A$ or to $\overline{A}$. The weights are
$\tilde w=2\cos(\pi e_0/2)$ and $w=2\cos(\pi e)$.

Relation~\eqref{eq:VV} allows us to derive the entanglement entropy $S(A)$
of segment $A$; see~\cref{fig:cft}(a). Each independent Bell cluster of the
PTIM that lives both in $A$ and $\overline{A}$ increases $S(A)$ by one. In
the picture of discrete bond percolation, such Bell clusters derive from
clusters of edges in space-time that connect $A$ with $\overline{A}$. Since
the loops of the continuum model essentially describe the \textit{boundaries}
of these clusters, we conclude that $S(A)\sim \langle N_A\rangle /2$.
The average number of loops $\langle N_A\rangle$ derives
from \eqref{eq:VV} by the relation $\langle N_A\rangle =\tilde
w[\partial_w\mathcal{V}_A(w)]_{w=\tilde w}$. Given the scaling dimension $h$
\eqref{eq:h} for the vertex operators, we find the logarithmic divergence
of the entanglement entropy
\begin{equation}
    S(A)
    \sim \frac{\langle N_A\rangle}{2}
    \sim \frac{\sqrt{3}\ln 2}{2\pi}\log_2 \Delta x\,.
\end{equation}
A comparison with Eq.~\eqref{eq:s_cft} implies the exact value of the
prefactor for the entanglement entropy at the critical point
\begin{equation}
    \tilde c=\frac{3 \sqrt{3}\ln 2}{2\pi}\,.
\end{equation}
It matches the numerical results in~\cref{fig:cft}(c) remarkably
well. This scaling has also been found in the Majorana representation of
Ref.~\cite{Nahum2020}.

Next, we focus on the mutual information $I(x_1,x_2)$ between the two
boundary points $x_1$ and $x_2$ of $A$; see~\cref{fig:cft}(b). For
$I(x_1,x_2)=1$, a percolation cluster that connects the sites $x_1$
and $x_2$ is required. This precludes clusters connecting the
interior of $A$ with the interior of $\overline{A}$. We therefore argue
that $I(x_1,x_2)\sim\langle \delta_{N_A=0}\rangle$. Note that we expect
$I(x_1,x_2)=2$ (indicating monogamous entanglement between the two sites) to
be irrelevant in the continuum limit. Using \eqref{eq:VV}, we find $\langle
\delta_{N_A=0}\rangle=\mathcal{V}_A(w \rightarrow 0)$. To set $w=2\cos(\pi
e)=0$, we choose $e=1/2$ and find with \eqref{eq:h} the scaling dimension
$h=1/3$ of the vertex operators. Consequently,
\begin{equation}
    I(x_1,x_2)\sim \langle\delta_{N_A=0}\rangle \sim \frac{1}{\left(\Delta x\right)^{\kappa}}
\end{equation}
with $\kappa=2h=2/3$, again consistent with numerical results to a remarkable
degree; see~\cref{fig:cft}(c).

\section{Relation to quantum error correction}
\label{sec:qec}

\begin{figure*}[tb]
    \centering
    \includegraphics[width=0.9\linewidth]{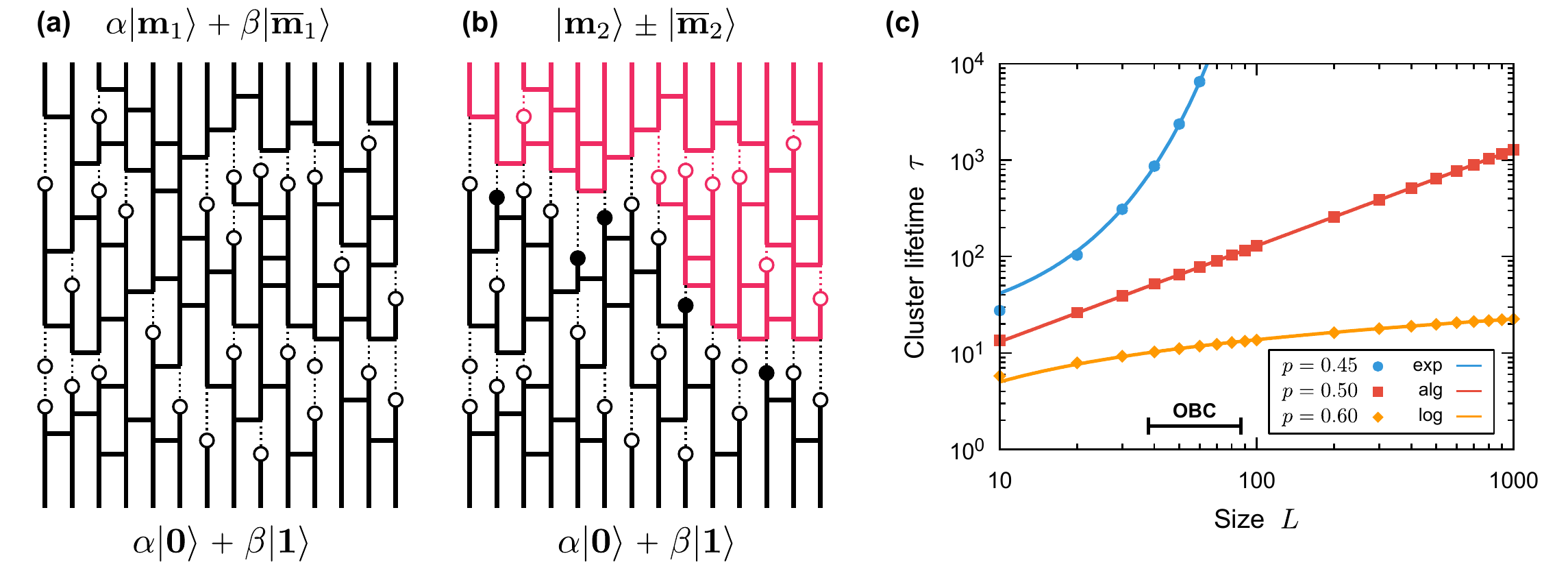}
    \caption{%
        \CaptionMark{Quantum error correction.}
        (a),(b)~Time evolution with random stabilizer measurements $S_e$
        between adjacent qubits (horizontal bars) and random errors $E_i$ on
        qubits (circles); time runs upwards. The system is initialized in
        the code space with a global Bell cluster (bold black vertical lines).
        (a)~The initial cluster survives and the amplitudes are preserved.
        (b)~Adding a few additional errors (black disks) makes the black
        cluster die out, losing all information about the encoded qubit. The
        new, independent cluster (red) is trivial and does not carry quantum
        information.
        (c)~Average cluster lifetime $\tau$ of a global Bell cluster as
        function of system size $L$ for $p\in\{0.45,0.50,0.60\}$. For
        $p=0.45<p_c$, $\tau$ grows exponentially so that the amplitudes of
        the initial cluster are retained almost indefinitely. At criticality
        $p=0.5=p_c$, $\tau$ grows algebraically with $\tau\sim L^\beta$ and
        $\beta\approx 1$. For $p=0.6>p_c$, the growth is only logarithmic. The
        solid lines are fits of the form $\alpha e^{\beta L}$ (blue), $\alpha
        L^\beta+\gamma$ (red), and $\alpha\log(L)+\beta$ (yellow). Each point is
        based on $10^5$ sampled trajectories on a chain with open boundary
        conditions and a cutoff simulation time $t_\mathrm{max}=5\times 10^4$.
    }
    \label{fig:QM}
\end{figure*}

Here we reinterpret the PTIM dynamics as a competition between projective
errors and syndrome measurements on a topological quantum memory---an approach
also successfully applied for the entanglement transition in the unitary
regime~\cite{Choi2019,Gullans2019a,Gullans2019b}. To this end, consider an
open chain of $L$ spinless fermions $c_i$ and define the Majorana modes
\begin{equation}
    \gamma_{2i-1}=c_i+c_i^\dag
    \quad\text{and}\quad
    \gamma_{2i}=i(c_i-c_i^\dag)
\end{equation}
with $\acom{\gamma_i}{\gamma_j}=2\delta_{ij}$, $\gamma_i^\dag=\gamma_i$,
and $\gamma_i^2=\mathds{1}$. Define stabilizer operators
$S_{e=(i,i+1)}=i\gamma_{2i}\gamma_{2i+1}$ for $i=1,\dots,L-1$, which obey
$S_e^\dag =S_e$, $S_e^2=\mathds{1}$ and $\com{S_e}{S_{e'}}=0$. We are
interested in the twofold degenerate ground state space of the quadratic
fermion Hamiltonian~\cite{Kitaev2001}
\begin{equation}
    H=-\sum_e S_e=-\sum_{i=1}^{L-1}\,i\gamma_{2i}\gamma_{2i+1}
    \label{eq:HMC}
\end{equation}
characterized by $S_e=1$ on all edges $e$. Let $\{\Ket{g_0},\Ket{g_1}\}$ be a
basis of the ground state space, the \emph{code space} of the Majorana chain
quantum code~\cite{Bravyi2010a}. A logical qubit with amplitudes $\alpha$
and $\beta$ is then encoded as $\Ket{\Phi}=\alpha\Ket{g_0}+\beta\Ket{g_1}$.

The elementary errors of the code are generated by the Hermitian on-site
operators $E_i=i\gamma_{2i-1}\gamma_{2i}$ and, due to $S_eE_i=-E_iS_e$
for $i\in e$, lead to excitations of the Hamiltonian~\eqref{eq:HMC}. In
the following, we assume that the environment measures $E_i$ projectively
with probability $p$ per site and time step. To detect and correct these
errors, we are allowed to measure the stabilizers $S_e$ projectively and use
the measurement outcomes, the so called \textit{error syndrome}. Common
schemes to protect the qubit $\Ket{\Phi}$ from decoherence employ
time-periodic measurements of all stabilizers and then use majority
voting on the syndromes to decide on unitary corrections in each time
step~\cite{Bravyi2010a,Lang2018}. Here we modify this scheme and perform
stabilizer measurements \emph{randomly} with probability $1-p$ per edge and
time step. This can be interpreted as a faulty implementation of the code
where syndrome measurements fail to be executed with probability $p$.

To reveal the connection to the PTIM, we need a spin-1/2 representation of
the fermion operators. We opt for the slightly unconventional Jordan-Wigner
transformation
\begin{equation}
    \gamma_{2i-1}=\prod_{j<i}\sigma_j^x\cdot \sigma_i^z
    \quad\text{and}\quad
    \gamma_{2i}=\prod_{j<i}\sigma_j^x\cdot \sigma_i^y
\end{equation}
which leads to the spin-1/2 representations
\begin{equation}
        S_e=\sigma_i^z\sigma_{i+1}^z
        \quad\text{and}\quad
        E_i=\sigma_i^x\,.
\end{equation}
An appropriate basis of the code space is then simply $\Ket{g_n}=\Ket{n\dots
n}$ with $\sigma_i^z\sigma_{i+1}^z\Ket{g_n}=\Ket{g_n}$ and $n=0,1$. In this
representation, the qubit is encoded as a global Bell cluster
\begin{equation}
    \Ket{\Phi}=\alpha \Ket{0\ldots 0}+\beta \Ket{1\ldots 1}\,,
    \label{eq:qubit}
\end{equation}
the random errors $E_i$ correspond to measurements of $\sigma_i^x$,
and the random measurements of stabilizers $S_e$ to measurements of
$\sigma_i^z\sigma_{i+1}^z$. We end up with a new interpretation of the PTIM
with open boundaries in one dimension, describing the competition between
errors and stabilizer measurements on a Majorana chain quantum code. This
representation has also been studied recently as an approximation for the
diffusion of Majorana defects~\cite{Nahum2020}.

In our previous studies of the PTIM, we initialized the system in the
unentangled product state $\Ket{\Psi(0)}=\Ket{+\dots+}$ and used the PTIM
to build up entanglement. Our discussion of the Majorana chain quantum
code suggests as initial state the global Bell cluster~\eqref{eq:qubit}.
How long does the system retain information about this qubit in the presence
of projective errors and random stabilizer measurements? In our analysis
of the PTIM dynamics, we found that the amplitudes $\alpha$ and $\beta$
survive the merging, growth, and shrinking of clusters. Therefore the
quantum information disappears irretrievably only if the initial cluster is
completely degraded. In~\cref{fig:QM} we sketch two quantum trajectories:
one where the initial cluster survives (a) and one where it decays (b).
For each quantum trajectory $\ket{\Phi(t)}$ with $\ket{\Phi(0)}=\ket{\Phi}$,
we can define the time $\tau_{\Phi}$ at which the initial cluster dies out.
We then define the average \emph{cluster lifetime} $\tau$ by averaging
$\tau_{\Phi}$ over many quantum trajectories. This time scale defines the
characteristic decay time of the stored quantum information in the system.

In~\cref{fig:QM}(c) we plot the scaling of this time scale $\tau$ with the
system size $L$ for different probabilities $p$. Again, the entanglement
transition is clearly visible at the critical value $p_c=0.5$: for $p<p_c$,
we find an exponentially diverging lifetime for increasing $L$, indicating
that quantum information can be robustly stored in large systems. By contrast,
numerics suggest that the growth of $\tau$ is only logarithmic for $p>p_c$,
while at criticality $p=p_c$ we find an algebraic behavior $\tau\sim L^\beta$
with $\beta\approx 1$. This observation relates the entanglement transition
to the capability of certain topological quantum memories to protect quantum
information from decoherence. More specifically, it provides an upper bound
for the Majorana chain quantum code with faulty stabilizer measurements,
above which the retrieval of quantum information is rendered impossible.
This relates to earlier findings by Aharonov \cite{Aharonov2000} who identified
a percolation-driven upper bound in generic, noisy quantum circuits above
which entanglement is necessarily short-ranged and scalable quantum computation
becomes impossible.

We now discuss the typical quantum trajectories in the two regimes in
detail. If we start with the initial state $\Ket{\Phi}=\alpha\Ket{\vec
0}+\beta\Ket{\vec 1}$, let the system evolve under conditions such
that the initial cluster survives [as in~\cref{fig:QM}(a)], and
finally measure the stabilizers on all edges, the system ends up in
the state $\Ket{\Phi'}=\alpha\Ket{\vec m}+\beta\Ket{\overline{\vec
m}}$ where again $\Ket{\overline{\vec m}}=\mathcal{U}\Ket{\vec m}$ with
$\mathcal{U}=\prod_i\sigma_i^x$. Note that, along the time evolution, the
sign between the two amplitudes of the initial cluster may change. However,
the last stabilizer measurements that condense all clusters into one
always reproduce the correct sign between the amplitudes [to see this, use
Eqs.~\eqref{eq:ex1}--\eqref{eq:ex4} and generalizations thereof]. The final
spin configuration $\vec m$ in $\ket{\Phi'}$ depends on the measurement
outcomes during the evolution and is only known if all measurements are
recorded. This implies that the quantum information is still stored in the
system---but to \emph{access} the qubit (mandatory for a useful quantum
memory), one has to deduce the configuration $\vec m$ from the collected
syndrome measurements in an efficient way. This decoding of the quantum memory
(and its efficiency) is beyond the scope of this paper.

Conversely, for quantum trajectories where the initial cluster dies
out and the quantum information is lost [as in~\cref{fig:QM}(b)], the
final state is $\Ket{\vec m}+\Ket{\overline{\vec m}}$ with probability
$|\alpha+\beta|^2/2$ and $\Ket{\vec m}-\Ket{\overline{\vec m}}$
with probability $|\alpha-\beta|^2/2$. Indeed, the error measurement
$\sigma_{i}^{x}$ that eventually removes the initial cluster determines
the sign and performs a ``measurement'' of the observable $\mathcal{U}$
on the stored qubit, projecting the system either into the symmetric or the
antisymmetric eigenspace of the symmetry $\mathcal{U}$.

\section{Summary and outlook}

In this paper, we introduced and studied the \emph{projective transverse field
Ising model}, a random circuit model where in each time step the noncommuting
observables $\sigma^x_i$ and $\sigma^z_i\sigma^z_{i+1}$ are measured randomly
with probabilities $p$ and $1-p$ on sites and edges, respectively. When
averaged over many quantum trajectories, the mutual information between far
apart spins behaves like a correlation function in conventional second-order
quantum phase transitions: while zero above a critical point $p_c$, it is
finite for $p<p_c$. This emergence of long-range entanglement between spins
is only visible in averages over quantum trajectories and not in the (up to
symmetries, maximally mixed) density matrix of the system. Using a classical
model for the entanglement dynamics---the \emph{colored cluster model}---we
performed extensive numerical simulations and presented an intuitive picture
of the entanglement transition which can be understood as the condensation
of colored clusters.

We would like to point out that this entanglement transition is not necessarily
linked to conventional phase transitions of nonequilibrium steady states
in driven dissipative systems~\cite{Diehl2008,Lang2015}. Indeed, if one
generalizes the projective transverse field Ising model to higher dimensions
and adds feedback to the process (say, spin flips that are conditioned on
the measurement outcomes), then long-range spin correlations are possible
and spontaneous symmetry breaking can occur in the steady state. Such
nonequilibrium phase transitions are reflected in the density matrix
and seem to be unrelated to the entanglement transition studied in this
paper. For instance, in one dimension, no long-range order is possible --
but the entanglement transition is still visible in ensembles of quantum
trajectories. In higher dimensions, both long-range order and the entanglement
transition can be observed; however, the critical points of these transitions
are not necessarily the same.

In a next step, we related the projective transverse field Ising model to
bond percolation on the space-time lattice of the process. This allowed us to
infer the critical point of the one-dimensional system exactly; we verified
this relation also for various lattices in two dimensions. Switching to the
continuum paved then the way for a conformal field theory of the critical
one-dimensional system. With this machinery, we derived the universal
prefactor $\tilde{c}=3\sqrt{3}\ln(2)/(2\pi)$ that describes the scaling
of the entanglement entropy, and the critical exponent $\kappa=2/3$ that
determines the algebraic decay of the mutual information. We compared these
results with numerical simulations and found almost perfect agreement with
the scaling behavior predicted by conformal field theory.

We concluded the paper with a discussion of the relevance of the entanglement
transition to quantum error correction by mapping the system to the Majorana
chain quantum code. In this context, the competing random measurements could
be identified as random stabilizer measurements and projective errors of
the environment. Contrary to typical scenarios, the outcomes of the syndrome
measurements were not employed for active error correction as our original
model did not include any kind of feedback. Despite this unconventional
setting, we showed that there is a hidden transition, parametrized by the
relative strength of stabilizer measurements and projective errors, that
separates two regimes: in one, the encoded amplitudes are quickly lost
irretrievably, whereas in the other, the lifetime of the amplitudes grows
exponentially with the system size. It is unclear whether this transition
is always identical to the well-known decoding thresholds of active quantum
error correction.


\begin{acknowledgments}
    We thank M.~P.~A.~Fisher for insightful discussions on entanglement
    transitions and M.~Ammon for discussions on aspects of conformal
    field theories. This research has received funding from the European
    Research Council (ERC) under the European Union's Horizon 2020 research
    and innovation programme (Grant agreement No.~681208). H.P.B.\ thanks the
    KITP for hospitality. This research was also supported by the National
    Science Foundation under Grant No.~NSF~PHY-1748958.
\end{acknowledgments}


\appendix

\section{Note on entanglement transitions}
\label{app:transition}

The entanglement transitions found in random quantum circuits (and
the one studied in this work) are characterized by the entanglement
of \emph{pure states} (quantified by entanglement entropy or mutual
information) when averaged over many independent quantum trajectories. For
a single quantum trajectory $\ket{\Psi(t)}$, the entanglement entropy of a
subsystem $A$ (at some fixed time) can be deduced from the density matrix
$\rho_A=\Tr{\ket{\Psi}\bra{\Psi}}{\ol A}$ which, in turn, is completely
determined by the expectation values of observables on $A$. In an experiment,
determining expectation values requires many copies of the same pure state
to average measurement outcomes. If the state in question \emph{could}
be prepared deterministically, this would only pose a technical but not a
fundamental problem. However, in the present context, the states of quantum
trajectories are prepared randomly with two sources of randomness. First,
entangling unitaries and projective measurements are \emph{chosen} randomly
according to some predefined distribution. As this randomness is of classical
origin (and therefore under the experimenter's control) it does \emph{not}
preclude the deterministic preparation of many copies of the same quantum
trajectory. However, the second source of randomness is due to the random
\emph{outcomes} of the projective measurements and therefore intrinsically
quantum. As we cannot apply projections deterministically, this makes the
preparation of many copies of a given quantum trajectory exponentially
unlikely and renders the experimental observation of such transitions elusive.

Mathematically, this issue can be traced back to the nonlinear dependence of
the entanglement entropy on expectation values of observables. To see this,
consider a quantity $\X$ that \emph{can} be written as a linear combination
of expectation values, $\X(\Ket{\Psi})=\sum_i\lambda_i\langle\O_i\rangle$;
then $\O_\X\equiv \sum_i\lambda_i\O_i$ is itself an observable such
that $\X(\Ket{\Psi})=\langle\O_\X\rangle$. We are interested in the
average of this quantity over many (say $M$) quantum trajectories
$\mathcal{N}=\{\Ket{\Psi(\bullet)}\}$ [recall Eq.~\eqref{sampleaverage}]
which---due to the assumed linearity---can be written as $X=\tr{\rho\,\O_\X}$
with density matrix
\begin{align*}
    \rho \equiv \frac{1}{M}\sum_{\Ket{\Psi}\in\mathcal{N}}\ket{\Psi}\bra{\Psi}\,.
\end{align*}
This quantity is experimentally accessible without any postselection as it
is merely an average of the observable $\O_\X$ over the ensemble $\rho$
of quantum trajectories. Unfortunately, the entanglement entropy depends
\emph{nonlinearly} on expectation values so that this argument breaks down.


%



\end{document}